\documentclass{Interspeech}

\interspeechcameraready 

\title{MOVER: Combining Multiple Meeting Recognition  Systems}

\author[affiliation={*,1}]{Naoyuki}{Kamo}
\author[affiliation={*,1}]{Tsubasa}{Ochiai}
\author[affiliation={1}]{Marc}{Delcroix}
\author[affiliation={1}]{Tomohiro}{Nakatani}

\affiliation{}{NTT Corporation}{Japan}

\email{naoyuki.kamo@ntt.com}

\usepackage{comment}
\usepackage{algorithm}
\usepackage{algpseudocode}
\usepackage{color}
\usepackage{multirow}
\usepackage{mathtools}
\usepackage{cite}

\algnewcommand{\Initialize}[1]{%
  \State \textbf{Initialize:}
  \State \hspace*{\algorithmicindent}\parbox[t]{0.8\linewidth}{\raggedright #1}
}
\begin{document}

\maketitle

\begin{abstract}
In this paper, we propose Meeting recognizer Output Voting Error Reduction (MOVER), a novel system combination method for meeting recognition tasks. Although there are methods to combine the output of diarization (e.g., DOVER) or automatic speech recognition (ASR) systems (e.g., ROVER), MOVER is the first approach that can combine the outputs of meeting recognition systems that differ in terms of both diarization and ASR. 
MOVER combines hypotheses with different time intervals and speaker labels through a five-stage process that includes speaker alignment, segment grouping, word and timing combination, etc.
Experimental results on the CHiME-8 DASR task and the multi-channel track of the NOTSOFAR-1 task demonstrate that MOVER can successfully combine multiple meeting recognition systems with diverse diarization and recognition outputs,
achieving relative tcpWER improvements of 9.55~\% and 8.51~\% over the state-of-the-art systems for both tasks.
\end{abstract}

\renewcommand{\thefootnote}{\fnsymbol{footnote}}
\footnote[0]{* Contributed equally}
\renewcommand{\thefootnote}{\arabic{footnote}}

\vspace{-3mm}
\section{Introduction}

With the advancements in deep learning technologies, the performance of automatic speech recognition (ASR) has remarkably improved \cite{prabhavalkar2023end,pmlr-v202-radford23a,haeb2020far}.
Thus, the research interests have moved to more challenging tasks such as meeting recognition \cite{kraaij2005ami,cornell23_chime,cornell24_chime,vinnikov24_interspeech,fu21b_interspeech,yu2022m2met}, which requires estimating \emph{who} speaks \emph{what} and \emph{when}, i.e., generating transcription with speaker labels and timing information for long multi-talker recordings including overlapped speech.

System combination is often used to boost meeting recognition performance \cite{yu2022summary,liang2023second,niu24_chime,mitrofanov24_chime,kamo24_chime,huang24b_chime,boeddeker23_chime,karafiat23_chime,deng23_chime}. For example, many of the CHiME-7 and 8 submissions used some kind of system combination, e.g., \cite{niu24_chime,mitrofanov24_chime,kamo24_chime,huang24b_chime,boeddeker23_chime,karafiat23_chime,deng23_chime}.
In these challenges, many participants used system combination with diarization-first pipelines \cite{raj2021integration,cornell23_chime,cornell24_chime}, which perform diarization, then separate each segment, and finally perform ASR on the separated signals. With such systems, we can combine the outputs of multiple diarization systems into a single diarization hypothesis using e.g., diarization output voting error reduction with overlap (DOVER-LAP) \cite{Raj2021Doverlap}.
We can then generate multiple recognition hypotheses for each segment of the diarization hypothesis using multiple ASR back-ends. Finally, we can combine the hypotheses for each segment using recognizer output voting error reduction (ROVER)~\cite{Fiscus1997ROVER}, minimum Bayes risk decoding \cite{XU2011802}, or N-best rescoring \cite{ROARK2007373}.

This approach can successfully reduce diarization and ASR errors.
However, we need to combine the diarization results into a single hypothesis to make sure that the different ASR back-ends generate hypotheses for the same speech content. This limitation prevents the correction of ASR errors due to the diarization (e.g., speaker confusion or missed detection) through the latter ASR system combination.

Other pipelines also exist for meeting recognition, such as the continuous speech separation (CSS) pipeline \cite{raj2021integration,9003827,vinnikov24_interspeech,10625894}, which performs separation, then transcribes the separated signals, and finally assigns speaker labels by clustering speaker embeddings for each word. Unlike diarization-first pipelines, it is not straightforward to combine the output of different CSS-based systems using, e.g., DOVER-LAP and ROVER, as there is no clear distinction between diarization and ASR steps.

Although it is well known that system combination is particularly effective when combining hypotheses with diverse error patterns \cite{zhou2012ensemble}, there are no approaches that can combine very distinct meeting recognition systems, such as diarization-first and CSS-based ones. To fully exploit such diversity, it is important to develop a system combination approach that can accommodate systems with different diarization and ASR outputs.

In this paper, we propose meeting recognizer output voting error reduction (MOVER) to combine the hypotheses of the multiple meeting recognition systems with different diarization and ASR outputs (i.e., transcriptions, timing information, and speaker labels).
Therefore, MOVER allows the combination of meeting recognition systems that could not be combined by conventional approaches.
Moreover, even for cases where we could use, e.g., DOVER-LAP and ROVER, we could improve the effect of the system combination because we can preserve the diversity of the hypotheses in terms of diarization and ASR.
The proposed MOVER algorithm can combine the hypotheses relying only on the meeting recognition outputs regardless of the system architecture, which allows us to combine even black box systems such as meeting recognition APIs.

We tested the effectiveness of the proposed MOVER on tasks 1 (DASR) and 2 (NOTSOFAR-1) of the CHiME-8 challenge.
We combined the meeting recognition systems of the different organizations 
with MOVER, and achieved relative tcpWER improvements of around 9\% 
compared to the best system of each task.
These results brought new state-of-the-art (SOTA) for these tasks, and
demonstrate the effectiveness of the proposed MOVER, showing that it can combine the meeting recognition systems with diverse diarization and ASR characteristics.
The scripts of the proposed MOVER are publicly released.\footnote{\url{https://github.com/nttcslab-sp/mover}}

\section{Related work}

The NIST Scoring Toolkit (SCTK)
\footnote{\url{https://github.com/usnistgov/SCTK}} provides the implementation of ROVER.
Although it basically assumes segmented transcriptions as input, it also has an optional mechanism that locates common silence regions among the hypotheses to segment long transcriptions~\cite{153216}.
This function requires the hypotheses with the start and end timing information for each word, but many meeting recognition systems only provide segment-level timing information.
Moreover, the tool supports only single-speaker recordings and does not handle the speaker ambiguities among hypotheses.
Thus, it is not possible to directly apply the tool to the meeting recognition tasks, which include long multi-speaker recordings with overlapped speech.

The authors in \cite{huang24b_chime} have mentioned they applied ROVER to merge meeting recognition systems with different diarization systems by matching the speaker labels of different systems and then performing ROVER on the whole text of each speaker.
However, the paper is mainly about the overall meeting recognition system description submitted to the CHiME-8 challenge\cite{huang24b_chime}. They did not describe the details of the system combination algorithm, e.g., how to merge the timing information of the segments or words.
To the best of our knowledge, this study is the first work describing a detailed algorithm extending ROVER for meeting recognition hypotheses and demonstrating its effectiveness through comprehensive experiments.

\section{Proposed method}

\subsection{System formulation}
\label{sec:system_formulation}

The aim of meeting recognition is transcribing a long recording with timing information and speaker labels for all segments in the recording (i.e., who speaks when and what?).
We assume that we have $H$ meeting recognition systems producing transcriptions with different errors.
Let $\mathbf{u}_{s,j}^{h} = (\mathbf{w}_{s,j}^{h}, \mathbf{v}_{s,j}^{h})$ be a pair of transcription $\mathbf{w}_{s,j}^{h}$ and segment timing $\mathbf{v}_{s,j}^h= [b_{s,j}^{h}, e_{s,j}^{h}]$ for the $j$-th segment of the speaker $s$ estimated by the meeting recognition system $h$, where $b_{s,j}^{h}$ and $ e_{s,j}^{h}$ are the start and end times of the segment.
Given a long recording, the meeting recognition system $h$ outputs the set of all segments of all speakers $\mathbf{O}^{h} =\{\mathbf{o}_{s}^{h}\}_{1\leq s \leq S^{h}}$, where $\mathbf{o}_{s}^{h}=\{\mathbf{u}_{s,j}^{h}\}_{1\leq j\leq J_{s}^{h}}$ denotes the set of all segments associated with the speaker $s$.
Here, $S^{h}$ is the total number of speakers, and $J_{s}^{h}$ is the number of segments of speaker $s$.

System combination aims at merging the output of the $H$ meeting recognition systems as: $\mathbf{O} = \{\mathbf{o}_{s}\}_{1 \leq s \leq S} =  \text{Combine}(\{\mathbf{O}^{h}\}_{1\leq h \leq H})$, where $\text{Combine}(\cdot)$ denotes the functional representation of the system combination, and $S$ is the number of speakers determined as a result of the combination.

\subsection{Conventional: ROVER}
\label{sec:conventional_rover}

ROVER is a classic system combination approach that merges the transcriptions of a segment estimated by multiple ASR systems.
The idea of ROVER is that each system has different error patterns.
Consequently, we may obtain more accurate transcriptions than any single system by combining multiple systems.
Let $\mathbf{w}^{h} = [w_{1}^{h}, w_{2}^{h}, \ldots, w_{N^{h}}^{h}]$ be the transcription (i.e., word sequence) estimated by the system $h$, where $N^{h}$ denotes its corresponding total number of words in the segment.
Since we consider only a single segment by a single speaker, we removed the $s$ and $j$ indices in this section.

ROVER mainly consists of two stages; 1) word alignment and 2) word voting.
First in the word alignment stage, by repeatedly applying a variant of dynamic programming (DP) matching~\cite{Sankoff1983}, we can obtain a confusion network (CN) $\{ (\hat{w}_{1}^{h=1}, \ldots, \hat{w}_{1}^{h=H}), \ldots, (\hat{w}_{N^{\text{CN}}}^{h=1}, \ldots, \hat{w}_{N^{\text{CN}}}^{h=H}) \}$, which represents the word alignment among the $H$ hypotheses.
Here, $N^{\text{CN}}$ is the number of aligned words.
$\hat{w}_{n}^{h}$ is the $n$-th word in the CN associated with hypothesis $h$, which corresponds to a word from the word sequence $\mathbf{w}^h$ or the \emph{null symbol}.
Then in the word voting stage, we can 
select the word according to some criterion, e.g., the word with the highest count, among all words in the same position $i$ and obtain the merged hypothesis $\mathbf{w}^{\text{ROV}} = \{w_{1}^{\text{ROV}}, w_{2}^{\text{ROV}}, \ldots, w_{N^{\text{ROV}}}^{\text{ROV}}\}$, where $N^{\text{ROV}}$ denotes the number of words determined as a result of voting.

\subsection{Proposed: MOVER}
\label{sec:proposed}

\begin{figure}[t]
\centering
\centering
 \includegraphics[width=0.99\linewidth]
 {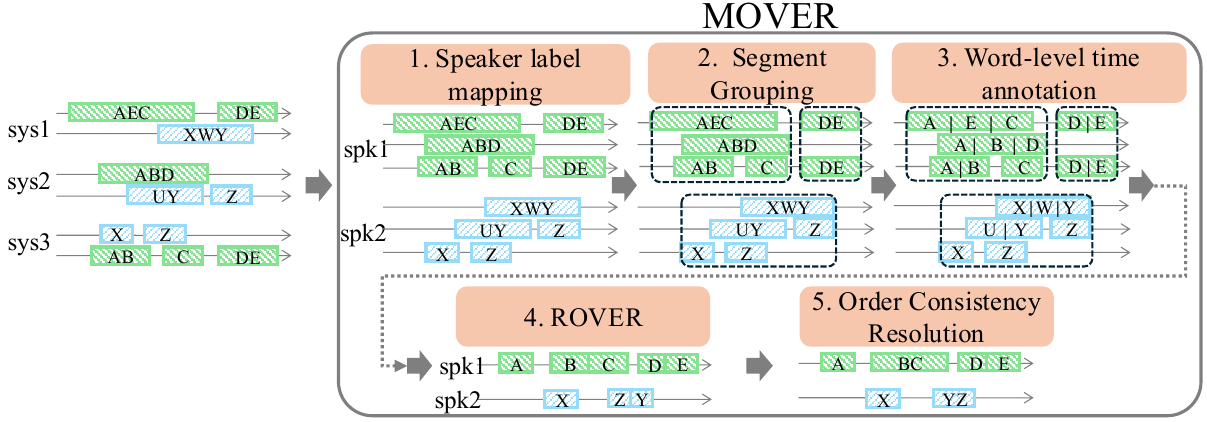} 
   \vspace{-1mm}
   \caption{Processing flow of MOVER}
   \label{fig:proposal}
   \vspace{0mm}
\end{figure}

As described in Section~\ref{sec:conventional_rover}, the ROVER algorithm assumes the transcriptions estimated by the multiple systems, which correspond to the same speech content with different errors.
When the systems share a diarization system (and use different ASR systems), the estimated segment boundaries are the same, and the transcriptions of each segment are aligned without any pre-processing.
However, the hypotheses of meeting recognition systems with different diarization systems can have a different number of segments and also different segment boundaries.
In addition, different meeting recognition systems may use different speaker label spaces; the order and number of speaker labels do not necessarily match across systems.
Consequently, to apply ROVER to the meeting recognition hypotheses, we need pre-processing to align the set of segments.

Figure~\ref{fig:proposal} shows the processing flow of the proposed MOVER, which comprises five stages; 1) speaker label mapping, 2) segment grouping, 3) word-level time annotation, 4) ROVER and timing combination, and 5) order consistency resolution (OCR).

{\bf 1) Speaker label mapping:} 
First, we align the transcriptions uttered by the same speaker across all systems.
To obtain such speaker-aligned hypotheses, we re-map the system-dependent speaker label space $s$ to the common label space $s^{\prime}$ using ${s^{\prime}} = \text{Map}(s, h)$, where $\text{Map}(\cdot)$ denotes the functional representation of the speaker label mapping.
If system $h$ has no output for speaker $s^{\prime}$,
we set $\mathbf{o}_{s^{\prime}}^{h} = \{ (\phi, \varphi) \}$, where $\phi$ and $\varphi$ denote the \emph{empty text and timing}, which is not judged to overlap with any segment. 

In this study, we adopted the global speaker mapping algorithm of the DOVER \cite{9004031}, which is based on the similarities between speaker time activities for the whole recording.

{\bf 2) Segment grouping:} 
In the second step, we group together transcriptions considered to be transcribing the same speech content.
We considered two approaches; 1) re-grouping the multiple segments and applying ROVER to each group (subset grouping) and 2) applying ROVER to the hypotheses concatenating all the segments over the entire recording (full set grouping).

The subset grouping divides the segments into groups as $\{ (\mathbf{g}_{s, k=1}^{h=1}, \ldots, \mathbf{g}_{s,k=1}^{h=H}), \ldots, (\mathbf{g}_{s,k=K}^{h=1}, \ldots, \mathbf{g}_{s,k=K}^{h=H}) \}$, where $K$ denotes the number of the re-grouped segments.
Here, $\mathbf{g}_{s,k}^{h} = \{ \mathbf{u}_{s,j=l_{k}^{h}}^{h}, \ldots, \mathbf{u}_{s,j=m_{k}^{h}}^{h} \}$ represents the re-grouped segment.
$l_{k}^{h}$ and $m_{k}^{h}$ denotes the start and end segment indices for the $k$-th re-grouped segment of the hypothesis $h$.

In our implementation, we group the segments over the hypotheses so that a segment in one group overlaps any segments in the group but does not overlap segments in other groups.
Algorithm~\ref{alg:segment_grouping} describes the procedure of the subset grouping, where $\Omega_{s}$ is the set of all indices $(h, j)$ of the hypotheses and segments for the speaker $s$, except empty hypotheses.
$\textbf{ARGSORT}(\cdot)$ returns the indices $(h, j)$ in ascending order of the start time of the segment.

On the other hand, the full set grouping simply makes a single group containing all the segments as $\{ (\hat{\mathbf{g}}_{s}^{h=1}, \ldots, \hat{\mathbf{g}}_{s}^{h=H}) \}$, where $\hat{\mathbf{g}}_{s}^{h} = \{ \mathbf{u}_{s,j=1}^{h}, \ldots, \mathbf{u}_{s,j=J_{s}^{h}}^{h} \}$.

\begin{algorithm}[t]
  \scriptsize
\caption{Segment grouping}
\begin{algorithmic}[0]
\Require{$ \{ (\mathbf{w}_{s,j}^{h}, [b_{s,j}^{h}, e_{s,j}^{h}]) \}_{(h,j) \in \Omega_{s}} $} \Comment{Set of segments}
\Ensure{$\tilde{\mathbf{G}}$} \Comment{Set of grouped segments} 
\State $\tilde{\mathbf{G}} \gets \phi$, $( \tilde{\mathbf{g}}^{1}, \ldots, \tilde{\mathbf{g}}^{H} ) \gets (\phi, \ldots, \phi)$, $\tilde{e} \gets 0$
\For{$h, j$ \textbf{in} $\textbf{ARGSORT}_{(h,j)}(\{{b}_{s,j}^{h}\}_{(h,j) \in \Omega_{s}})$} \Comment{Sort by start time}
\If{$ (\tilde{e} = 0)$ \textbf{or} $(b_{s,j}^{h} < \tilde{e})$}  \Comment{If empty-sets or overlapped}
    \State $\tilde{\mathbf{g}}^{h} \gets \tilde{\mathbf{g}}^{h} \cup \{ (\mathbf{w}_{s,j}^{h}, [b_{s,j}^{h}, e_{s,j}^{h}]) \}$
    \State $\tilde{e} \gets \text{max}(\tilde{e}, e_{s,j}^{h})$ \Comment{Get the maximum end time of segments}
\Else  \Comment{If not overlapped}
    \State $\tilde{\mathbf{G}} \gets \tilde{\mathbf{G}} \cup \{ (\tilde{\mathbf{g}}^{1}, \ldots, \tilde{\mathbf{g}}^{H}) \}$ \Comment{Appent the group}
    \State $(\tilde{\mathbf{g}}^{1}, \ldots, \tilde{\mathbf{g}}^{H}) \gets 
    (\phi, \ldots, \phi), \tilde{e} \gets 0$ \Comment{Initialize}
\EndIf
\EndFor
\end{algorithmic}
\label{alg:segment_grouping}
\end{algorithm}

{\bf 3) Word-level time annotation:} MOVER utilizes the word-level timing information for the ROVER-based merging procedure of the transcription and timing information.
Meeting recognition systems do not always provide word-level timing (only segment-level timing).
Thus, when the hypotheses do not have word-level time information, we require assigning it by, e.g., applying forced alignment~\cite{mcauliffe17_interspeech}, pseudo-word-level annotation \cite{MeetEval23}, etc.
The segment-level timing information $[b_{s,j}^{h}, e_{s,j}^{h}]$ is converted to word-level one $[b_{s,j,n=1}^{h}, e_{s,j,n=1}^{h}], \ldots, [b_{s,j,n=N_{s,j}^{h}}^{h}, e_{s,j,n=N_{s,j}^{h}}^{h}]$, where $b_{s,j,n}^{h}$ and $e_{s,j,n}^{h}$ are the start and end times of the $n$-th word of the $j$-th segment, and $N_{s,j}^{h}$ is the number of words in it.

In this study, we adopted the pseudo-word-level annotations~\cite{MeetEval23} with \textit{character-based} option for each segment, which assigns the time interval for each word in proportion to its number of characters.

{\bf 4) ROVER and timing combination:} Finally, given the set of grouped segments $(\mathbf{g}_{s,k}^{h=1}, \ldots, \mathbf{g}_{s,k}^{h=H})$, we apply ROVER to each group of segments and obtain the merged meeting recognition hypothesis $\mathbf{g}_{s,k}^{\text{ROV}} = (\mathbf{w}_{s,k}^{\text{ROV}}, \mathbf{v}_{s,k}^{\text{ROV}})$.
By applying DP matching to the hypotheses, we can obtain the aligned words $(\hat{w}_{s,j,n}^{h=1}, \ldots, \hat{w}_{s,j,n}^{h=H})$ and their timing information $([b_{s,j,n}^{h=1}, e_{s,j,n}^{h=1}], \ldots, [b_{s,j,n}^{h=H}, e_{s,j,n}^{h=H}])$.
Then, we obtain the merged transcription $\mathbf{w}_{s,k}^{\text{ROV}}$ by the word voting as for ROVER (see Section \ref{sec:conventional_rover}).
In our implementation, we simply select the most frequent word for voting.
When the null symbol is selected by the voting, it is removed from the merged hypothesis.

We also merge the timing information for each word as $b_{s,j,n}^{\text{ROV}} = \frac{1}{N_{s,j,n}}\sum_{h \in \mathcal{H}_{s,j,n}} b_{s,j,n}^{h}$ and $e_{s,j,n}^{\text{ROV}} = \frac{1}{N_{s,j,n}}\sum_{h \in \mathcal{H}_{s,j,n}} e_{s,j,n}^{h}$, where $\mathcal{H}_{s,j,n}$ is the set of hypotheses that have the same word as that selected by the voting process and $N_{s,j,n}$ is the number of hypotheses in $\mathcal{H}_{s,j,n}$.

As a result of the segment grouping, the duration of the re-grouped segment may become very long (especially, for the full set grouping), which could cause the word alignments whose word timings are temporally distant and may cause a negative impact on the system combination.
To prevent such implausible alignments, we proposed to extend the DP matching procedure of ROVER's word alignment stage by incorporating the time-constraint mechanism (TC-DP), like tcpWER \cite{MeetEval23}.
Unlike the DP for a pair of hypothesis and reference transcriptions in the tcpWER, ROVER requires performing the DP between the CN (aligned hypotheses) and a hypothesis.
When we compute the score (i.e., \emph{correct}, \emph{substitution}, \emph{insertion}, and \emph{deletion}) between a word in the hypothesis and one of the words in the CN, we apply a time constraint, allowing \emph{correct} and \emph{substitution} only if the word timings overlap.
To allow for a certain amount of time gaps between hypotheses, we adopted a collar $c$ like \cite{MeetEval23}, which we set to $5$ seconds based on our preliminary experiments.

{\bf 5) Order consistency resolution:}
The simple merging procedure of the timing information can cause a mismatch between the order of the words and their timing information, which may change the order of the words from that obtained with ROVER.
Here, we hypothesize that the word order of ROVER is more accurate and propose a post-processing procedure called OCR to correct the timing information.

Algorithm~\ref{alg:ocr} describes the OCR procedure, where $\textbf{delete}$ stands for removing an element from a list. 
The idea of OCR is to find order inconsistent words that have an end time after the start time of the next word, concatenate these words, and re-assign the timing information of the concatenated words.
In this study, we applied the OCR to the merged hypothesis 
$\{ (\mathbf{w}_{s,j=1}^{\text{ROV}}, \mathbf{v}_{s,j=1}^{\text{ROV}}), \ldots, (\mathbf{w}_{s,j=J}^{\text{ROV}}, \mathbf{v}_{s,j=J}^{\text{ROV}}) \}$ for each speaker $s$, where $J$ is the total number of the segments in the entire hypothesis.
Each of the element $(\mathbf{w}_{s,j}, \mathbf{v}_{s,j})$ contains only one word $\mathbf{w}_{s,j} = [w_{s,j}]$ and its word-level timing information $\mathbf{v}_{s,j} = [b_{s,j}, e_{s,j}]$.
Here, we assume that the words are arranged in the order of ROVER's output, and $w_{s,j}$ is the $j$-th word in the entire transcription.

\begin{algorithm}[t]
  \scriptsize
\caption{Order consistency resolution}
\begin{algorithmic}[0]
\Require{$ \{ (\mathbf{w}_{s,j}, [b_{s,j}, e_{s,j}]) \}_{1 \leq j \leq J} $ } \Comment{Set of segments of merged hypothesis}
\Ensure{$\{ (\tilde{\mathbf{w}}_{j}, [\tilde{b}_{j}, \tilde{e}_{j}]) \}_{1 \leq j \leq \tilde{J}} $} \Comment{Set of revised segments}
    \State $ \{(\tilde{\mathbf{w}}_{j}, [\tilde{b}_{j}, \tilde{e}_{j}]) \}_{1 \leq j \leq J}  \gets  \{(\mathbf{w}_{s,j}, [b_{s,j}, e_{s,j}]) \}_{1 \leq j \leq J} $ \Comment{Copy input}
    \State $j \gets 1, \tilde{J} \gets J$
    \While{$j < \tilde{J}$}
        \If{$\tilde{b}_{j+1} < \tilde{e}_{j}$} \Comment{If order is inconsistent or overlapped}
        \State $[\tilde{b}, \tilde{e}] \gets [\text{min}(\tilde{b}_{j},~ \tilde{b}_{j+1}), \text{max}(\tilde{e}_{j},~ \tilde{e}_{j+1})]$  \Comment{Merge two segments}
        \State $\tilde{\mathbf{w}} \gets [\tilde{\mathbf{w}}_{j}, \tilde{\mathbf{w}}_{j+1}]$ \Comment{Merge two segments}
        \State $(\tilde{\mathbf{w}}_{j}, [\tilde{b}_{j}, \tilde{e}_{j}]) \gets (\tilde{\mathbf{w}}, [\tilde{b},\tilde{e}])$ \Comment{Replace old to new element}
        \State $\textbf{delete}~~ (\tilde{\mathbf{w}}_{j+1}, [\tilde{b}_{j+1}, \tilde{e}_{j+1}])$ \Comment{Remove old element from set} 
        \State $j \gets 1, \tilde{J} \gets \tilde{J} - 1$ \Comment{Go to start of the loop}
        \Else
        \State $ j \gets j + 1$
        \EndIf 
    \EndWhile
\end{algorithmic}
\label{alg:ocr}
\end{algorithm}

\section{Experiments}
\subsection{Experimental setup}

We evaluated the effectiveness of the proposed MOVER on task-1 (DASR) and the multi-channel track of task-2  (NOTSOFAR/mc) of the recent CHiME-8 Challenge.
The evaluation set of task 1 comprises four datasets, i.e., CHiME-6~\cite{chime6}, DiPCO~\cite{Segbroeck2019}, Mixer 6~\cite{brandschain2010mixer}, and NOTSOFAR-1~\cite{vinnikov24_interspeech}, while that of the Task 2 focuses on the NOTSOFAR-1 dataset.
Following CHiME-8's regulation, we evaluated the performance using the tcpWER with a collar of 5 seconds.
All texts in the hypotheses were normalized using the CHiME-8 text normalization tool
, which includes unifying numerical representations, removing fillers, etc \cite{cornell24_chime}.

For our experiments, we used the system outputs provided by the participants. For task 1, we combined systems from the two participating teams (STCON~\cite{mitrofanov24_chime} and NTT~\cite{kamo24_chime}), and two additional post-challenge systems \cite{kamo2025microphonearraygeometryindependent}.
For task 2, we used the systems from four teams, i.e., USTC-NERCSLIP~\cite{niu24_chime},  STCON~\cite{mitrofanov24_chime},  NTT~\cite{kamo24_chime}, and NPU-TEA~\cite{huang24b_chime}.

\vspace{-1mm}
\subsection{Experimental result of Task 2 (NOTSOFAR/mc)}

\begin{table}[t]
    \centering
    \caption{Comparison of the tcpWER [\%] on the eval set of NOTSOFAR/mc for the 1-best system and different configurations of MOVER for the combination of nine systems.
    }
    \vspace{-2mm}
    \resizebox{0.7\linewidth}{!}{
    \begin{tabular}{ccccccc}
        \hline
        System & Grouping & OCR & TC-DP & Eval \\
        \hline
        1-Best & - & - & - & 10.81 \\
        \hline
        \multirow{6}{*}{MOVER} &\multirow{3}{*}{Full set} & & & 11.29 \\
                               & & \checkmark & & 10.26 \\
                               & & \checkmark & \checkmark & \textbf{9.89} \\
        \cline{2-5}
                               &\multirow{3}{*}{Subset} & & & 10.81 \\
                               & & \checkmark & & 9.95 \\
                               & & \checkmark & \checkmark & 9.96 \\         
        \hline
    \end{tabular}
    }
    \vspace{0mm}
    \label{tab:notsofar1}
\end{table}

\begin{figure}[t]
\centering
\centering
 \includegraphics[width=0.75\linewidth]{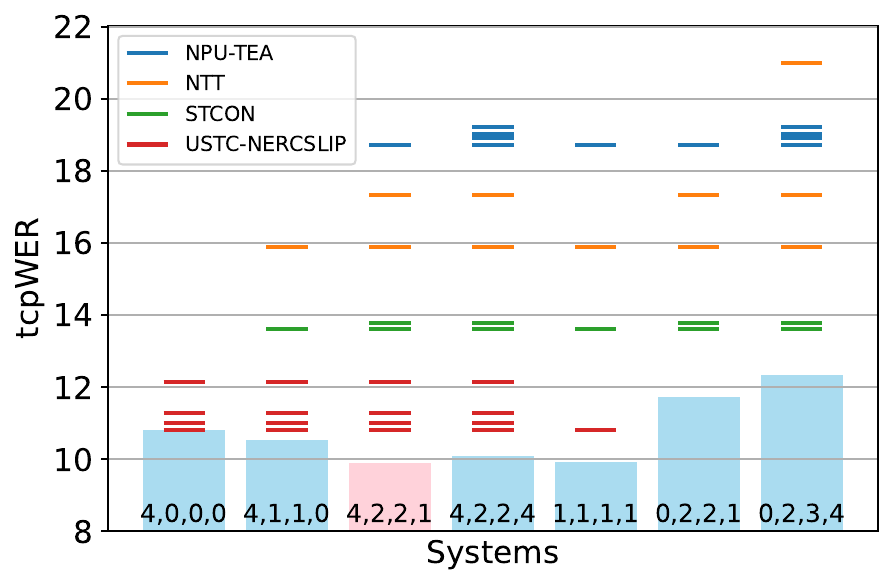}
   \vspace{-3mm}
   \caption{tcpWER [\%] when combining different systems with MOVER. 
   Each bar represents the performance of the combined system (the pink one is the best). The horizontal lines show the performance of the individual systems. The numbers at the bottom indicate the number of systems from each team (i.e., from left to right:  USTC-NERCSLIP, STCON, NTT, and NPU-TEA).}
   \label{fig:bar}
   \vspace{0mm}
\end{figure}

Table~\ref{tab:notsofar1} shows the results of the  NOTSOFAR/mc task.
We selected nine systems among all CHiME-8 submissions for combination, based on the performance on the development set; four systems from USTC-NERCSLIP~\cite{niu24_chime}, two systems from STCON~\cite{mitrofanov24_chime}, two systems from NTT~\cite{kamo24_chime}, 
and one system from NPU-TEA~\cite{huang24b_chime}. 
We evaluated the three options of MOVER, i.e., 
Full set or Subset ``Grouping'', without or with ``TC-DP'', and without or with ``OCR''.
As a baseline, the table includes the  USTC-NERCSLIP/sys1, which achieved the best performance on the evaluation set using a single system (1-Best).

From Table \ref{tab:notsofar1}, we observe that MOVER can improve performance over the 1-best system for both full set and subset grouping. The ablation results show that resolving the consistency of the word orders with OCR is essential for MOVER to outperform the 1-Best system. TC-DP does not improve the performance for the subset grouping, while it is essential for the full set grouping. This result is reasonable, because the DP matching for the long segment may cause implausible word alignments with a large time discrepancy.
Overall, MOVER+TC-DP+OCR with Full set grouping achieved the best performance and a 0.9-point improvement compared to the baseline.\footnote{We also evaluated system combination with SCTK ROVER using speaker label mapping and pseudo-word-level annotation. It failed to improve performance over the 1-best, achieving performance comparable to that of MOVER with full set grouping without OCR and TC-DP (11.32 vs. 11.29 \%).}
The difference between 1-Best and the best MOVER approaches statistical significance according to a one-sided Student’s t-test (p = 0.057).\footnote{The tcpWER is evaluated for each session, and thus the number of samples is small (i.e., 160 samples), which may explain why the test did not reach a smaller p-value.}
In the following, we report only the results of this configuration.

Note that the performance after applying MOVER (i.e., 9.89~\%) is the new SOTA performance of the NOTSOFAR-1/mc task, outperforming the top challenge system (i.e., 10.81~\%, NERCSLIP/sys1). These results demonstrate that the proposed MOVER can successfully merge diverse meeting recognition systems, designed by different organizations.

Figure~\ref{fig:bar} shows the results obtained by combining different systems with MOVER.
The results show that MOVER stably improved performance for various combinations of the systems. 
Depending on the combined systems, the performance improvement varies. 
For example, combining the 4 best systems did not improve performance, probably because of the lack of variety, as they were all built by the same team (USTC-NERCSLIP). The best performance was achieved by combining nine systems from four teams, but interestingly, a very close performance could be achieved by combining only each team's best system.

These results suggest that it is more effective to merge systems with diverse diarization and ASR characteristics than to merge systems with good performance but similar characteristics.
These results re-confirm the potential of MOVER as it allows merging systems with different diarization results, resulting in more diverse ASR results.

\vspace{-1mm}
\subsection{Experimental result of Task 1 (DASR)}
We evaluated the effectiveness of MOVER on four different datasets of task 1 (DASR) of CHiME-8. Only two teams (STCON and NTT) submitted a total of five systems. In addition, we included the output of two more systems built after the challenge
~\cite{kamo2025microphonearraygeometryindependent}, bringing the total to seven systems.

Table~\ref{tab:dasr_eval} compares the best-performing systems of STCON and NTT with the results obtained with MOVER.
MOVER improves the macro-averaged tcpWER by 1.1 points over the SOTA (i.e., 19.6 \%, STCON/sys2).
Except for Mixer-6, MOVER stably achieved the performance gains over the best-performing single systems.
These results confirm the effectiveness of the proposed method not only for NOTSOFAR-1 but also for other meeting recognition datasets.
The STCON and NTT systems utilized system combination methods for both diarization and ASR outputs.
Even for such systems, the proposed MOVER obtained further performance gains.

\begin{table}[t]
    \centering
    \caption{tcpWER [\%] results on the eval set of the DASR task.}
    \vspace{-2mm}
    \resizebox{1.0\linewidth}{!}{
    \begin{tabular}{@{}lccccc@{}}
        \hline
         & CHiME-6 & DiPCo & Mixer 6 & NOTSOFAR-1 & Macro-Ave \\
        \hline
        NTT/best & 31.99	&21.68&	13.74	&13.83	&20.31 \\ 
        STCON/sys2 & 33.13	&19.88&	\textbf{10.87}	&14.61	&19.62 \\ 
        \hline
        MOVER    &  \textbf{30.32}&	\textbf{18.61}&	12.28&	\textbf{12.97}&	\textbf{18.54}\\
        \hline
    \end{tabular}
    }
    \vspace{-1mm}
    \label{tab:dasr_eval}
\end{table}

\vspace{-1mm}
\section{Conclusion}

In this paper, we proposed a novel system combination method for meeting recognition, called MOVER.
MOVER is a voting-based method that combines multiple hypotheses with different transcriptions, timing information, and speaker labels.
With the proposed MOVER, we could combine systems from different CHiME-8 participants, achieving new SOTA performance on both the DASR and NOTSOFAR/mc tasks, demonstrating its effectiveness. 
%
The proposed implementation of MOVER uses global speaker mapping.
In future work, we would like to explore using local speaker mapping to allow further speaker confusion error correction based on textual information.


\clearpage

\bibliographystyle{IEEEtran}
\bibliography{mybib}

\end{document}